# Probing the geological setting of exoplanets through atmospheric analysis: using Mars as a test case

Monica Rainer[a], Evandro Balbi[b,c], Francesco Borsa[a], Paola Cianfarra[b], Avet Harutyunyan[d], Silvano Tosi[a,c,e]

[a]INAF - Osservatorio Astronomico di Brera, Via E. Bianchi, 46, Merate (LC), 23807, Italy

[b]University of Genoa, Department of Earth, Environment and Life Sciences, Corso Europa, 26, Genoa, 16132, Italy

[c]INFN - Sezione di Genova, Via Dodecaneso 33, Genoa, 16146, Italy

[d]INAF - Fundación Galileo Galilei, Rambla José Ana Fernandez Pérez 7, Breña Baja (TF), 38712, Spain

[e]University of Genoa, Department of Physics, Via Dodecaneso 33, Genoa, 16146, Italy

**Abstract**

One of the frontier research fields of exoplanetary science is the study of the composition and variability of exoplanetary atmospheres. This field is now moving from the gas giant planets towards the smaller and colder telluric planets, and future instruments like ANDES will focus on the observations of the atmosphere of telluric planets in the habitable zone in reflected light. These future observations will possibly find variable signals due to the view of different hemispheres of the planet. Particularly, the strength of the signal may be linked to the thickness of the atmospheric layer probed, and therefore to the average altitude variations of the planetary surface, that are related to the global geodynamic evolution of the planet. To better prepare for the interpretation and exploitation of these future data, we used Mars as a Solar System analog of a spatially resolved telluric exoplanet. We observed the reflected light of Mars with the high-resolution near-infrared (NIR) spectrograph GIANO-B (widely used in exoplanetary atmospheric studies) during a 3 month period: we studied the spatial and temporal variations of the Martian $CO_2$ signal using the least-squared deconvolution technique (LSD), to mimic as closely as possible the standard exoplanetary atmospheric analysis. We linked the variations found to the well-known Martian geological surface characteristics: we found a clear dependence of the strength of the $CO_2$ signal with the thickness of the Martian atmospheric layer by comparing the retrieved $CO_2$ signal with the altitudes of our pointings. The proposed strategy is promising: it proved to be effective on Mars and may shed light on the variations in the strength of atmospheric signal of telluric exoplanets.

1. **Introduction**

Since the discovery of the first exoplanet orbiting a solar-like star (Mayor and Queloz, 1995), the field has continuously expanded, now also dealing with atmospheric investigation and characterization. The atmospheric studies are still mostly focused on Hot and Ultra-Hot Jupiters (HJs and UHJs): the former are defined by their very short periods (<10 days, Wang et al., 2015), i.e. their proximity to their host star, while the latter are a subset of the HJs with an equilibrium temperature larger than approximately 2200 K (Parmentier et al., 2018).

Recent spectroscopic observations of exoplanetary atmospheres at high-resolution have shown hints of variability in UHJs, using both transmission and emission spectroscopy techniques (e.g., Nugroho et al., 2020; Rainer et al., 2021; Borsa et al., 2021b, 2022; Giacobbe et al., 2021). However, whether these variabilities are caused by true atmospheric conditions or by other factors (e.g. the analysis technique used, the telluric contamination removal, the often low signal-to-noise ratio of the extracted planetary spectrum, or the underestimation of the errors) is still debated. Unfortunately, there is no planet with similar characteristics in our Solar System that we could use as an analog to guide the interpretation of exoplanetary observations. Recently, signals attributed to possible atmospheric variations have been observed in the rocky super-Earth 55 Cancri e (Patel et al., 2024; Loftus at al., 2025), but, while being a telluric planet, this is a ultra-short-period (USP) magma world with a highly variable dayside temperature that may reach up to 2700 K (Demory et al., 2016).

New instruments and the refinement of the analysis' techniques are moving the focus towards the smaller and colder telluric planets that are well represented in our Solar System. The astronomical community is building the new generation of high-resolution optical and near-infrared (NIR) spectrographs for the 40-m class telescopes, e.g. ANDES[1] for the ELT. One of the main scientific cases of ANDES will be the observation of telluric planets in reflected light (Marconi et al., 2022).

We could then exploit the possibility to observe a very well-known telluric planet such as Mars and spatially resolve its atmospheric signal to link any eventual variability to known surface properties. This is a great advantage over the study of a telluric exoplanet: the exoplanet cannot be spatially resolved, and as such any recovered signal would depend only on the global properties of the visible hemisphere. Using the detailed geological characteristics of Mars as a lead would help to interpret the wealth of atmospheric exoplanet data that will arrive in the near future also in relation to the more global exoplanetary geological characteristics. We focused on the study of the Martian $CO_2$ taking advantage of the very strong signal expected. Furthermore, there is no $CO_2$ in the solar spectrum (Asplund et al., 2021), which helps disentangling the Martian atmospheric spectrum from the reflected solar one.

Low-resolution measurements of the Martian $CO_2$ column density have been used in the past to build topographical maps of the planet (Bibring et al., 1991), so we planned to observe several different regions of Mars and then link the equivalent width (EW) of the $CO_2$ cross-correlated profile of each observational pointing to the average altitude of the pointing region, which roughly anti-correlate with the atmospheric thickness. We also wanted to perform variability studies of the Martian atmosphere on a time-basis of a few months to see if we could derive information on the seasonal changes of the atmosphere itself, if differential measurements of the two hemispheres could be conducted.

---

[1] `https://elt.eso.org/instrument/ANDES/`

Mars may not be used as a proxy for all kinds of temperate or cold telluric planets, as its atmospheric thickness and composition, and its surface conditions may be unique, but it is still the best candidate to start our investigation. The thick atmosphere and cloud cover of Venus would hinder any topographical analysis, while the rocky and icy moons of Jupiter or Saturn (with Titan being the more interesting target for atmospheric analysis) may not be spatially resolved from ground, so we could only link our data to the characteristics of a whole hemisphere. Still, we plan to apply the same strategy described in this paper to other molecules and other telluric bodies of the Solar System in the future, to see if it would be possible to discriminate between Martian-like atmospheric variations, driven by the geological setting, and other situations, as the Venus-like clouds.

We studied the variations of the Martian atmospheric $CO_2$ signal using high-resolution NIR spectroscopy and analysis techniques derived from the field of exoplanetary atmospheric studies. The two goals of our work were: i) to link any characteristics of the atmospheric signal to the geological properties of the surface, serving as a basis for the future characterization of telluric exoplanets in reflected light; and ii) to conduct a variability monitoring and look for seasonal variations of the atmospheric signal. Unfortunately, the second goal was unattainable due to adverse weather conditions that limited our temporal monitoring, so we focus here on using the very well-known information of the Martian topography to link the atmospheric signals retrieved with exoplanetary analysis' techniques to the geological setting of a telluric planet.

We detail our observational strategy, the dataset and the data reduction in Sect. 2, with particular attention to the telluric contamination removal process. The data analysis and the variability measurements are described in Sect. 3, and our results are reported in Sect. 4. Finally, our conclusions and plans for future works are in Sect. 5.

## 2. Observations and data reduction

We observed Mars using the high-resolution NIR slit spectrograph GIANO-B (Oliva et al., 2012; Origlia et al., 2014; Claudi et al., 2018) at the Telescopio Nazionale Galileo (TNG) in the framework of the observing program A46TAC_22 (PI Rainer). GIANO-B covers the spectral range 950-2450 nm with a nominal resolving power R ~ 50,000. The offered observing modes are the nodding mode (for point-like objects) and the stare mode (for extended objects): we observed Mars in stare mode.

We optimized our observational strategy to monitor possible temporal variations of the Martian atmospheric signal. The Martian northern hemisphere's spring equinox was on 26 December 2022, so we planned six visits distributed before, around, and after this date to observe the southern summer-autumn or the northern winter-spring transition. Additionally, we planned to observe Mars during a period close to its opposition phase, to exploit its large apparent angular diameter: GIANO-B slit has a size of 6 x 0.5 arcsec (length x width), so to ensure that we could always observe the two hemispheres independently regardless of the slit inclination on the planetary disk (a variable depending only on instrumental constraints that we could not control) we needed Mars to have an apparent angular diameter larger than 12 arcsec, i.e. at least twice the slit's length. Each visit consisted of six spatially diverse pointings, with each pointing composed of five short 10 seconds exposures (to avoid saturation) that we could then sum up to increase the signal-to-noise ratio (SNR), as can be seen in Table 1. Unfortunately, the weather conditions at the TNG during autumn and winter 2022 were adverse. Thus, only three visits were completed (2022-10-09, 2022-11-17, and 2022-12-16), and although the time span approaching the equinox was homogeneously covered, we were not able to conduct any observation after that date. This prevented us from monitoring any eventual seasonal variation.

## 2.1. Pointings

With the aim of pointing at specific regions of the Martian surface, we did not use the GIANO-B autoguider, that would automatically set the slit on the center of the observed object, but we used the guiding camera to manually observe a chosen position. The position of the slit at the beginning and at the end of the five 10 seconds exposures was registered, so we could estimate any possible drift for each pointing. Figure 1 shows the position of the slit for a single pointing as seen by the observer at the telescope: although the details of the surface are slightly blurred, they are clear enough to allow the identification of the different Martian regions and global-scale physiographic features (e.g., Tharsis, Thaumasia, Valles Marineris - see Table 1 for details). We were able confirm the stability of the pointings, which is mostly comparable to the slit's width of 0.5 arcsec, and well below the slit's length of 6 arcsec. This allowed us to sum up the five exposures to increase the SNR. Appendix A shows all the pointings and the slit drifts for the three visits we performed on Mars. Table 1 lists the regions that we explored in our three visits, along with a note on the slit's drift. We identified the pointing regions by comparing the images of Mars obtained during our observations (Figs. A.16, A.17, and A.18) with the global albedo map obtained with the Thermal Emission Spectrometer instrument (TES, Christensen et al., 2001) aboard the Mars Global Surveyor spacecraft (MGS): see Figs. A.13, A.14, and A.15. We note here that due to the relative dimensions of the slit and the apparent angular diameter of Mars during our visits, the slit covers regions ~2500-3000 km long depending on the night.

| Pointing | Region | Coordinates | Drift [arcsec] | Altitude [m] | Alt. St.Dev. [m] | SNR [H band] |
|---|---|---|---|---|---|---|
| 2022-10-09 ||||||| 
| 1 | South Polar Cap | 70° S 30° W | 0.96 | 1389 | 316 | 1171 |
| 2 | Argyre Planitia, Noachis Terra | 40° S 45° W | 0.28 | 1002 | 1110 | 1092 |
| 3 | Valles Marineris, Xante Terra, Arabia Terra | 0° N 45° W | 0.83 | -466 | 2070 | 1250 |
| 4 | Lunae Planitia, Chryse Planitia, Arabia Terra | 25° N 35° W | 0.68 | -3127 | 816 | 1216 |
| 5 | Tharsis Montes, Lunae Planum | 10° N 85° W | 0.57 | 1108 | 3456 | 1166 |
| 6 | Tempe Terra, Acidalia Planitia | 40° N 50° W | 0.74 | -3238 | 1816 | 1023 |
| 2022-11-17 |||||||
| 1 | Arabia Terra, Terra Sabaea | 10° N 25° E | 0.81 | -597 | 623 | 967 |
| 2 | Noachis Terra, Terra Sabaea | 25° S 25° E | 0.45 | 1596 | 441 | 800 |
| 3 | Noachis Terra | 55° S 20° E | 0.76 | 792 | 1210 | 795 |
| 4 | Acidalia Planitia | 45° N 5° E | 0.33 | -4236 | 344 | 728 |
| 5 | Chryse Planitia, Arabia Terra | 30° N 15° W | 0.19 | -3211 | 721 | 889 |
| 6 | Noachis Terra | 40° S 5° W | 1.48 | 1182 | 671 | 737 |
| 2022-12-16 |||||||
| 1 | Aonia Terra | 60° S 80° W | 3.18 | 960 | 1137 | 1063 |
| 2 | Claritas Fossae, Thaumasia Highlands | 35° S 80° W | 0.31 | 3469 | 1410 | 1082 |
| 3 | Tharsis Montes, Lunae Planum | 5° N 80° W | 1.60 | 1634 | 2054 | 1430 |
| 4 | Lunae Planum, Chryse Planitia, Arabia Terra | 25° N 45° W | 0.34 | -2544 | 1557 | 1224 |
| 5 | Olympus Mons, Tharsis | 15° N 115° W | 0.62 | 2470 | 3267 | 1438 |
| 6 | Alba Patera | 45° N 110° W | 0.37 | 345 | 1827 | 1310 |

Table 1: Number of pointings, region identification along with the central coordinates, slit's drift, altitude (average and standard deviation along the slit), and SNR of the combined spectrum in the H band for each pointing of the three visits on Mars. The average altitude along the slit's position is referred to the Martian geoid. The slit covers a large portion of the Martian surface, therefore some of the pointings cover a very large range of altitudes and subsequently show a large standard deviation.

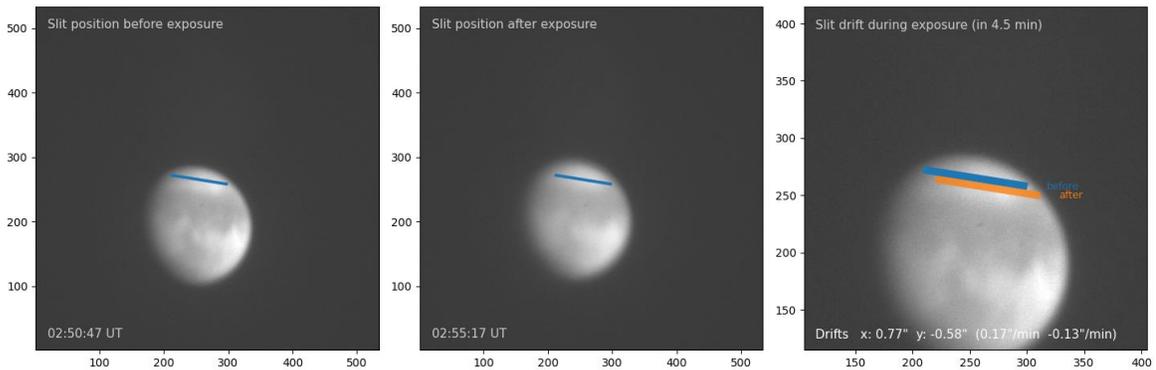

Figure 1: Pointing on the Martian southern polar region. The slit drift before (left panel) and after (middle panel) the sequence of five exposures is quite small, as shown in the right panel.

The average altitude of each pointing was derived in the Geographic Information System environment (GIS) using the freely available QGIS software[2] by tracing a topographic profile along the position of the slit. Topographic data were derived from the `MGS MOLA - MEX HRSC Blended DEM Global v2` (Fergason et al., 2018) that assembles the Digital Elevation Model (DEM) derived from the Mars Orbiter Laser Altimeter (MOLA) instrument belonging to the NASA's MGS, and the High-Resolution Stereo Camera (HRSC), belonging to the ESA's Mars Express (MEX) with spatial resolution of 200 m/px. We sampled the topographic profile inside the slit in 1000 steps, which allowed us to recover the average altitudes and their standard deviation along the profiles (see Table 1).

---
[2] v3.18Zurich, https://www.qgis.org/it/site/forusers/download.html

### 2.2. Data reduction

GIANO-B is equipped with a online data reduction software (DRS) (Harutyunyan et al., 2018). A stand-alone part of the DRS is the freely available[3] software `Gofio` (Rainer et al., 2018), that performs all the standard reduction steps (dark and flat-field subtraction, cosmic rays removal, wavelength calibration), after the ramp-processing (i.e., starting from the raw FITS data). `Gofio` outputs two final products for each observed spectrum: i) the usual wavelength calibrated, barycentric corrected monodimensional spectrum with the echelle orders merged and a constant step in wavelength (the so-called `s1d` spectra), and ii) a wavelength calibrated, not barycentric corrected spectrum with the echelle orders separated and the original constant step in pixel (the so-called `ms1d` spectra). In this work, we used the latter, less manipulated spectra, where the flux is not interpolated on a constant step grid, and there is no pseudo-normalisation performed in order to merge the echelle orders.

### 2.3. Telluric contamination removal

The contamination of the telluric bands is very strong in the NIR wavelength range. In our case, the telluric component was also highly blended with the Martian one due to the relative radial velocity (RV) of the planet (see Sect. 3): a precise and careful removal of the telluric contamination was fundamental for our work. Thus, we decided to use the ESO software

---

[3] `https://atreides.tng.iac.es/monica.rainer/gofio`

`molecfit v4.3.1`[4] (Smette et al., 2015; Kausch et al., 2015) to remove the telluric contamination from our spectra.

GIANO-B is an echelle spectrograph, which means that each echelle order is independently wavelength calibrated (Oliva et al., 2018). The U-Ne lamp used for this calibration does not work equally well in all orders, resulting in an inhomogeneous precision in the different echelle orders' wavelength calibration. An incorrect wavelength calibration may cause the appearance of p-Cygni profiles during the telluric removal, due to a misalignment between the observed spectra and the computed telluric transmission spectrum: thus, we decided to correct each single order by itself so that `molecfit` could compensate for the slight differences in the wavelength precision in the various orders. Using `esoreflex`, the graphical interface of `molecfit`, we defined for each order the molecules to remove and the regions to use to compute the atmospheric model (well positioned along the whole order, uncontaminated by scientific features).

We were particularly interested in clearing the spectral regions around the Martian $CO_2$ features. To avoid removing the Martian features, we used `molecfit` to model another molecule present in the same orders (typically $H_2O$), and then `molecfit` adjusted the abundance of telluric $CO_2$ using the MIPAS[5] and GDAS[6] atmospheric profiles to scale the value relative to the computed molecular abundance. The wavelength solution from the modeled molecule is also used for the whole computed telluric transmission spectrum. Figure 2 shows the results in a region rich in $CO_2$ lines, where both the telluric and Martian individual $CO_2$ lines are visible. The Martian lines are much deeper: even if the Martian atmosphere is much thinner than the terrestrial one, with an atmospheric volume less than 1% of Earth's (Haberle, 2015), still the higher concentration of $CO_2$ (~95% of the total atmosphere, Franz et al., 2017) in comparison to the terrestrial one (~0.0417% of the total atmosphere at the time of the observations[7]) results in a very strong Martian absorption. We note here that we used only a subset of GIANO-B orders (23 out of 50) for our analysis, discarding those completely absorbed by telluric water bands, or with several individual but very strong telluric absorption lines where `molecfit` may sometimes leave an emission residual, and those with none or very faint $CO_2$ lines.

---

[4] `https://www.eso.org/sci/software/pipelines/skytools/molecfit`
[5] `https://earth.esa.int/eogateway/instruments/mipas`
[6] `https://www.ncei.noaa.gov/products/weather-climate-models/global-data-assimilation`

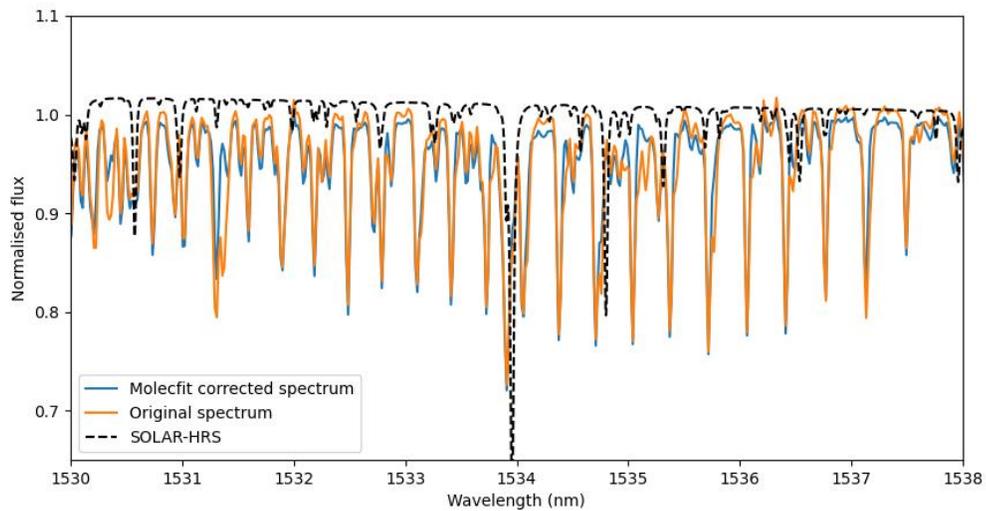

Figure 2: Selected wavelength region of a GIANO-B spectrum where many $CO_2$ absorption lines are found. The original spectrum (orange line) is composed both by the telluric and the Martian $CO_2$ lines, with the Martian signal being much stronger due to the high concentration of $CO_2$ in the Martian atmosphere. The telluric-corrected spectrum is shown in blue: only the Martian $CO_2$ remains. The dashed black line shows the solar SOLAR-HRS spectrum in the same wavelength range: even if there is no $CO_2$ in the solar atmosphere, still some other absorption lines are present in the same wavelength range and they are blended with the Martian spectrum. The slight shift between the solar and the GIANO-B spectra is due to the different RV of the solar spectrum as reflected by the Martian surface in the GIANO-B spectrum.

---

[7] `https://climate.nasa.gov/vital-signs/carbon-dioxide/`

## 3. Data analysis

Our aim was detecting spatial and temporal variations, both in strength and dynamics, of the Martian $CO_2$ signal. Thus, we applied a variant of the cross-correlation technique to enhance our signal and recover the mean line profile of the Martian $CO_2$. This was done also to mimic the techniques used in the spectroscopic high-resolution studies of exoplanetary atmospheres, even if there is a crucial difference between our data and the standard exoplanet observations in reflected light: in the exoplanetary spectra are present the atmospheric signal, the stellar reflected signal, and the telluric contamination, but also the stellar spectrum itself, as the whole star-planet system is observed at the same time. In our data, instead, the stellar spectrum itself is not present, since we are in the observational condition of putting only the planet within our slit, without the direct presence of the host-star (i.e., the Sun in this case) inside the slit.

We remind here briefly of the more common workflow used in the spectroscopic high-resolution studies of the exoplanetary atmospheres (Birkby, 2018), that takes advantage of the geometry of a system consisting of an host star and a transiting exoplanet: during the transit it is possible to observe, e.g., the transmission spectrum of the planet, while the thermal emission spectrum is observed near the secondary eclipse of the system (e.g., Pino et al., 2020; Borsa et al., 2022). Even in the case of non-transiting planets, it is possible to observe the thermal emission if the geometry of the star-planet system allow us to see the planet's day-side (e.g., Brogi et al., 2012). In the same conditions, also the stellar light reflected by the planet's surface will be observed, and if the exoplanet is cold enough only the reflected light will be visible (e.g., Birkby et al., 2017; Strachan and Anglada-Escude, 2020; Di Marcantonio et al., 2019). In all cases, it is a common strategy to observe a continuous spectroscopic time-series that covers not only the transit time or the time before and/or after the secondary eclipse (or the orbital phases when the day-side is visible), but also an almost equivalent period before/after transit or during the secondary eclipse (or when the day-side is not visible). In this way, a series of purely stellar spectra is recovered, and an average stellar spectrum is computed and removed from all the other star-planet blended spectra. Still, in the cases of the emission spectroscopy and the observations of exoplanets in reflected light, a stellar component will remain in the stellar light reflected by the planetary surface.

Once the non-reflected stellar spectrum is removed, the planetary spectra are moved in the planet's rest-frame exploiting the possibility given by HR spectroscopy of resolving the star and planet reference systems (Birkby, 2018), and then combined to enhance the atmospheric signal. Some very strong absorption lines (e.g., H, Na, Mg) may be observed directly (Wyttenbach et al., 2015; Yan and Henning, 2018; Borsa et al., 2021a), while for many elements or molecules the planetary signal is too faint, and the cross-correlation technique is implemented (Snellen et al., 2010). In these cases, a model of the expected absorption spectrum of a single element/molecule at the conditions of temperature and pressure of the planet is built, possibly using one of the many tools available to the community (e.g. Molliere et al., 2019; Kitzmann et al., 2023), and the observed spectrum is cross-correlated with the model, to recover an enhanced signal of the desired element. The more absorption lines are present both in the spectrum and in the model, the stronger the recovered signal will be, as long as the position of the model lines is precise, to avoid a smearing of the resulting cross-correlation function (CCF).

We note here also that, while the studies of exoplanetary atmospheres with high-resolution spectroscopy at the beginning focused only on the optical range, they expanded in recent years to exploit also the NIR wavelength range. GIANO-B in particular has been widely used, and its wide wavelength range allowed the detection of numerous molecules (Giacobbe et al., 2021; Guilluy et al., 2022; Carleo et al., 2022), so we are working with an instrument already well-known in the field. Still, we were not able to follow all the steps described above in our work, because there are some crucial differences between the observation of a exoplanet star system and a single Solar System body. First of all, we could observe Mars individually, so that the removal of the independent solar spectrum is not needed, similarly to how ANDES will observe exoplanets in reflected light using the IFU mode with adaptive optics (Palle et al., 2025).

Additionally, the Martian signal is very strong, so we did not need a long spectroscopic time-series to enhance our signal: while we combined the spectra of each pointing, we did not create a single planetary spectrum for each visit. This was done also to exploit the geological information of the Martian surface and to consider each pointing as the global view of an exoplanetary hemisphere with specific characteristics. We could then consider each single pointing as an exoplanetary spectrum observed in reflected light, after the removal of the independent stellar spectrum and the combination of the planetary time-series.

We built a $CO_2$ template covering the wavelength range 1000-2300 nm starting from a transmission spectrum simulated with `petitRADTRANS` (Molliere et al., 2019), assuming an isothermal atmospheric profile with T=210 K (which is the average Martian temperature) and a surface pressure level of 10 mbar (the average Martian surface pressure). Starting from this template, which was used to identify the laboratory wavelengths of the absorption lines of $CO_2$, we then created a binary mask using the positions of the maxima. To avoid giving too much weight to small lines, we excluded all the lines fainter than 10% with respect to the largest one in the wavelength interval, ending with a total of 2410 lines.

The Martian spectra were affected not only by the tellurics contamination, but also by the reflected solar spectrum, which falls almost at the same RV of the Martian absorption spectra: we recovered the RVs of the two components (Martian and solar) querying the JPL Horizons service[8] with the `astroquery` package[9] (see Table 2). While there are no $CO_2$ lines in the solar spectrum, there may be other absorption lines that fall in the same wavelength regions. The disk-integrated high-resolution solar spectrum (SOLAR-HRS[10]) from Meftah et al. (2023) is plotted in Figure 2 with a dashed black line: to avoid solar contamination we took out from our $CO_2$ mask all the spectral lines distant less than 15 km s$^{-1}$ from any solar line. This decreased the number of lines of the $CO_2$ mask from 2410 to 1710 lines. Then we built our solar mask by selecting all the local minima of the SOLAR-HRS spectrum in the same wavelength range of the $CO_2$ mask (1000-2300 nm), and discarding all the lines contaminated by the Martian $CO_2$. We ended up with 2311 lines in the solar mask.

---

[8] `https://ssd.jpl.nasa.gov/horizons/`

[9] `https://astroquery.readthedocs.io/en/latest/`

[10] `http://bdap.ipsl.fr/voscat_en/solarspectra.html`

Once we had all the relevant masks, we used a self-written `python` package to perform a least-squares deconvolution (LSD) between our observed spectra and the masks, following the numerical method of Donati et al. (1997). We chose the LSD technique over the more standard cross-correlation because, while numerically very similar, the LSD method reduce the distortion due to line blending and the presence of regular patterns in the lines' positions (Kochukhov et al., 2010). The drawback of this method is a usually lower SNR of the resulting mean line profile, but in our case this was not an issue, as we were working with very high SNR spectra. We computed our profiles in the velocity range from -40 km s$^{-1}$ to +40 km s$^{-1}$, using a step of 0.2 km s$^{-1}$.

| Pointing | RV Mars [km s$^{-1}$] | RV Sun-Mars [km s$^{-1}$] | RV Sun [km s$^{-1}$] |
|---|---|---|---|
| 2022-10-09 | | | |
| 1 | -9.940 | 2.095 | -7.845 |
| 2 | -9.924 | 2.095 | -7.823 |
| 3 | -9.914 | 2.095 | -7.819 |
| 4 | -9.903 | 2.095 | -7.808 |
| 5 | -9.893 | 2.095 | -7.798 |
| 6 | -9.882 | 2.095 | -7.787 |
| 2022-11-17 | | | |
| 1 | -4.392 | 2.265 | -2.127 |
| 2 | -4.382 | 2.265 | -2.117 |
| 3 | -4.372 | 2.265 | -2.107 |
| 4 | -4.359 | 2.265 | -2.094 |
| 5 | -4.350 | 2.265 | -2.085 |
| 6 | -4.340 | 2.265 | -2.075 |
| 2022-12-16 | | | |
| 1 | 5.310 | 2.197 | 7.507 |
| 2 | 5.322 | 2.197 | 7.519 |
| 3 | 5.336 | 2.197 | 7.533 |
| 4 | 5.353 | 2.197 | 7.549 |
| 5 | 5.365 | 2.197 | 7.561 |
| 6 | 5.377 | 2.197 | 7.574 |

Table 2: Radial velocities of Mars with respect to the observing location on Earth (TNG), Sun with respect to Mars, and reflected solar light with respect to the TNG.

By comparing the original and telluric corrected spectra, we can see the goodness of our telluric removal despite the heavy blending of the Martian and telluric $CO_2$ feature: the upper panels of Fig. 3 show the comparison between the $CO_2$ profiles of the first pointing of each visit obtained using the original and the `molecfit` corrected spectra. Because our spectra are still not corrected for the barycentric velocity, the telluric feature falls always at 0 km s$^{-1}$. The solar profiles obtained with the SOLAR-HRS mask for the same spectra are also shown in the lower panel of Fig. 3: comparing them with the Martian $CO_2$ profiles, is it clearly seen that the two components have indeed almost the same RV.

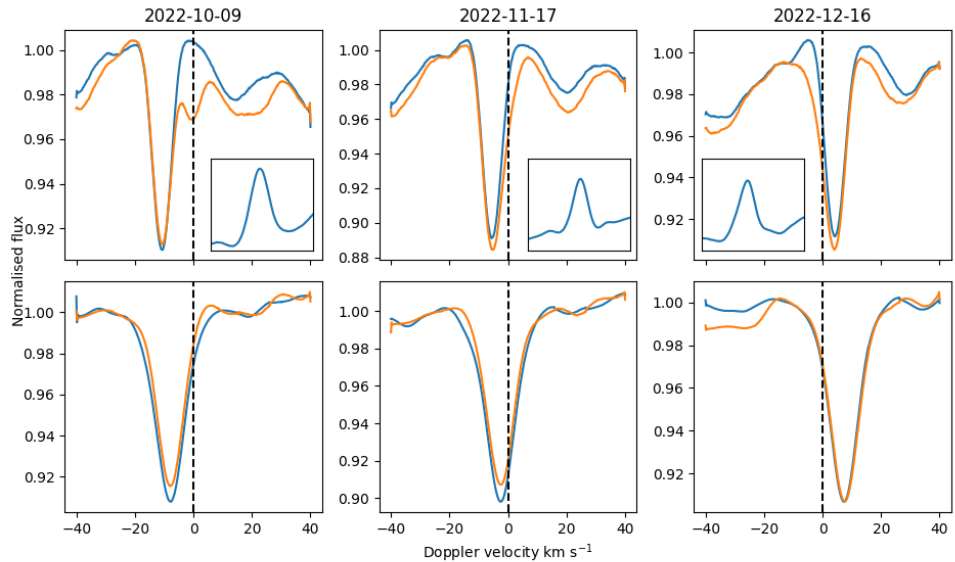

Figure 3: Martian $CO_2$ and reflected solar light mean line profiles. *Upper panel:* comparison between the $CO_2$ profiles before (orange line) and after (blue line) the telluric removal. The dashed vertical lines show where the telluric signal is expected to fall, i.e. at 0 km s$^{-1}$. The examples are representative of the situation during the three visits: the blending increases due to the different velocity of Mars relative to Earth. The inserts show the telluric $CO_2$ signal, recovered by subtracting the telluric-affected profile from the cleaned one. All the inserts cover the [-20,20] km s$^{-1}$ range on the x-axis and from 0 to 0.04 in normalised flux on the y-axis. The distortions on the line continuum are due to both the fringing that affects the GIANO-B spectra and the correlation with the $CO_2$ lines that we excluded from our mask due to solar contamination. *Lower panel:* profiles of the solar reflected spectra. In this case, the telluric contamination is very faint, due to the absence of the telluric molecules in the solar spectrum.

### 3.1. Radial velocities of the $CO_2$ profiles

High-resolution spectroscopy studies of exoplanetary atmospheres often use very stable optical spectrographs, that are specifically optimized for RV measurements. As such, the sign of possible atmospheric variability can also be seen in the variations in the RV of the atmospheric signal (e.g. Rainer et al., 2021). Unfortunately, the wavelength calibration of GIANO-B may suffer from the sparse useful emission lines of the calibrating U-Ne lamp in some echelle orders, hindering its capability to reach the same RV precision as its optical counterparts, regardless of its stability.

Still, in this study we are looking only for RV variations, so we shouldn't suffer for any eventual zero point error, and as we completed each visit in about one hour of observing time we could disregard any instrumental drift in such a short time (we remind here that is not possible to observe a scientific target with a simultaneous wavelength calibrating lamp with GIANO-B). With these assumptions, we computed the RVs of our spectra for each pointing and then we looked at variations inside each single visit.

We derived the RVs from our mean line profiles by normalising them first and then fitting them with a Gaussian function using the `curve_fit` function of the `python` module `scipy`[11]. The fit returns both the central position of the line, i.e. our RV measurement, and its error, taken from the co-variance matrix of the fit (see Table 3). While a Voigt profile would be a more faithful representation of the absorption line, we decided here to follow the standard methodology used in exoplanetary studies, that computes the RVs using a Gaussian fit of the CCF (Queloz, 1995).

Of course, the RVs are in the terrestrial rest-frame, because we computed our profiles from the `ms1d` spectra, where the barycentric correction is not applied. To bring the RVs to their true values we had to take into account the motion of Mars relative to the Earth (for the Martian $CO_2$ profiles), and the relative velocity of the Sun with respect to Mars (for the solar profiles). As stated before, we used the `astroquery` package to query the JPL Horizons System and recover the velocities information for Mars at the time of our observations relative to our observing position (TNG at the Roque de Los Muchachos Observatory). The results are listed in Table 2. We could then recover the true RV values of our mean line profiles, that are listed in the fourth and fifth columns of Table 3 (RV and $RV_{Sun}$).

The RVs show a consistent deviation from the zero value, which may be due to a physical effect (Martian rotation, presence of winds in the atmosphere), or to the slightly problematic wavelength calibration of the GIANO-B spectra. We decided to exploit the simultaneous solar reflected spectrum present in our spectra as a reference to compensate for any cause different from true atmospheric movements. The rotation velocity of the Martian regions probed by our pointings, the wavelength calibration of GIANO-B, or to any instrumental effects should affect both set of data in the same way. This is actually the case, as it is shown in Fig. 4.

---

[11] https://docs.scipy.org/doc/scipy/

While the Sun has its own radial velocity variations, due to stellar activity phenomena or p-mode oscillations, their expected amplitude is very low, of the order of the m s$^{-1}$ for the former (Lanza et al., 2016), and cm s$^{-1}$ for the latter (Kjeldsen and Bedding, 2011). Both values are well below the errors on our RVs, as such we felt confident in using the solar RVs to correct the Martian values. For each visit, we computed the linear correlation between solar and Martian RVs, then we used it to correct the Martians RVs. The original RVs, the solar RVs and the resulting corrected Martian RVs are listed in Table 3. The variations of the final RVs are shown in Fig. 5: while mostly consistent with a zero value there are some clear deviations.

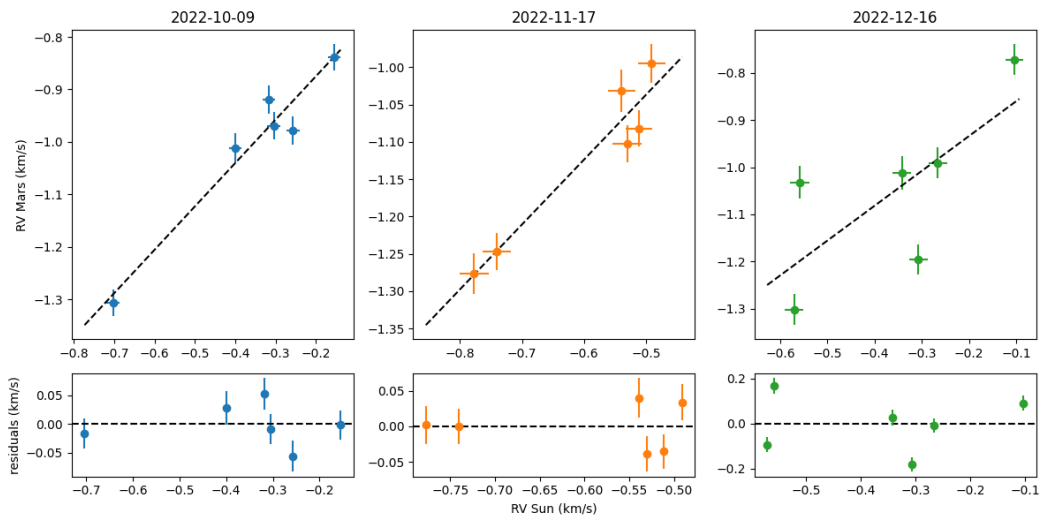

Figure 4: Comparison between the RVs of the Martian $CO_2$ and those of the solar profiles: each column represents a visit. *Upper panels:* direct comparison between the solar (x-axis) and Martian (y-axis) RVs. The linear correlation between the two sets of data is shown as a dashed black line. *Lower panels:* residuals of the Martian RVs after removing the correlation with the solar values.

| Pointing | MJD [days] | EW [km s$^{-1}$] | RV [km s$^{-1}$] | RV$_{Sun}$ [km s$^{-1}$] | RV$_{corrected}$ [km s$^{-1}$] |
|---|---|---|---|---|---|
| 2022-10-09 | | | | | |
| 1 | 59862.11948 | 0.672±0.017 | -0.839±0.025 | -0.155±0.016 | -0.001±0.025 |
| 2 | 59862.12792 | 0.670±0.018 | -0.978±0.027 | -0.257±0.016 | -0.056±0.027 |
| 3 | 59862.13299 | 0.709±0.019 | -0.969±0.026 | -0.304±0.016 | -0.008±0.026 |
| 4 | 59862.13833 | 0.746±0.020 | -0.919±0.027 | -0.317±0.016 | 0.053±0.027 |
| 5 | 59862.14300 | 0.745±0.019 | -1.307±0.026 | -0.703±0.017 | -0.016±0.026 |
| 6 | 59862.14811 | 0.742±0.021 | -1.012±0.029 | -0.400±0.016 | 0.029±0.029 |
| 2022-11-27 | | | | | |
| 1 | 59900.97121 | 0.831±0.021 | -0.995±0.026 | -0.491±0.022 | 0.034±0.26 |
| 2 | 59900.97670 | 0.704±0.020 | -1.031±0.028 | -0.539±0.022 | 0.040±0.028 |
| 3 | 59900.98247 | 0.850±0.021 | -1.103±0.025 | -0.530±0.023 | -0.039±0.025 |
| 4 | 59900.98994 | 0.853±0.022 | -1.082±0.024 | -0.511±0.021 | -0.036±0.024 |
| 5 | 59900.99435 | 0.882±0.022 | -1.246±0.025 | -0.741±0.023 | -0.001±0.025 |
| 6 | 59900.99979 | 0.776±0.022 | -1.276±0.027 | -0.777±0.023 | 0.002±0.027 |
| 2022-12-16 | | | | | |
| 1 | 59929.92558 | 0.689±0.023 | -0.991±0.033 | -0.266±0.020 | -0.008±0.033 |
| 2 | 59929.93082 | 0.651±0.021 | -1.195±0.032 | -0.307±0.020 | -0.182±0.032 |
| 3 | 59929.93659 | 0.613±0.022 | -1.012±0.035 | -0.342±0.020 | 0.027±0.035 |
| 4 | 59929.94330 | 0.764±0.025 | -0.772±0.033 | -0.103±0.019 | 0.091±0.033 |
| 5 | 59929.94815 | 0.680±0.023 | -1.302±0.033 | -0.571±0.020 | -0.094±0.033 |
| 6 | 59929.95298 | 0.666±0.023 | -1.032±0.035 | -0.559±0.020 | 0.167±0.035 |

Table 3: Results from the mean line profile analysis: for each pointing, the Modified Julian Date (MJD) at mid-exposure, the EW of the Martian $CO_2$ profile, the RVs of the Martian and solar profiles, and the resulting corrected Martian RV are listed.

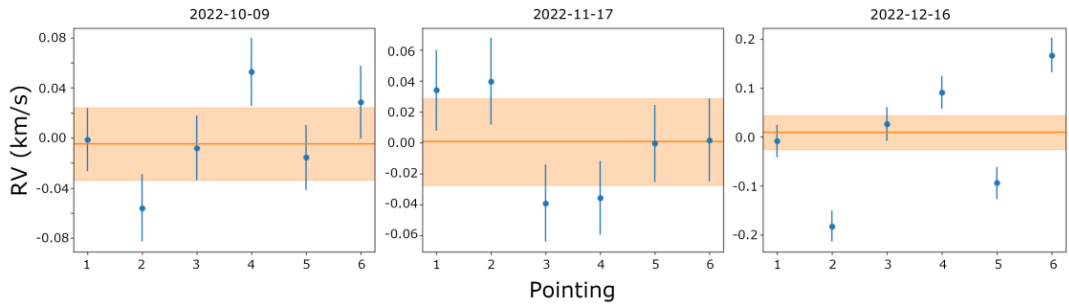

Figure 5: Spatial variations of RV of the Martian $CO_2$ profiles. The orange line is the median value, the shaded orange area shows the region around the median within the maximum error computed on the measured RVs.

The final RVs range from -182 to 167 m s$^{-1}$ (see last column of Table 3, and Fig. 6 to see the RV values on an Aitoff projection of Mars), with a median value of 0±71 m s$^{-1}$. If we consider the three nights independently, we still find that the mean RV values are consistent with zero: $RV_{mean}$ = 0 ± 34 m s$^{-1}$ for 2022-10-09, $RV_{mean}$ = 0 ± 30 m s$^{-1}$ for 2022-11-17, and $RV_{mean}$ = 0 ± 115 m s$^{-1}$ for 2022-12-16. We searched for correlations between the RVs and the weather condition during each visit (airmass, humidity, pressure, wind speed, temperature), but we found none (the Spearman correlation between RVs and the above values yielded p-values ranging from 0.30 and 0.92), so we can assume that the variations found are real and neither of telluric origin nor an artifact of the reduction process, which is the same for each spectrum.

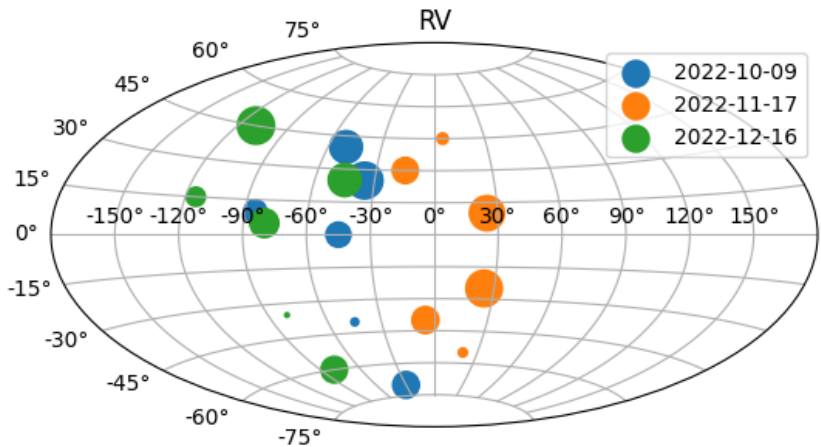

Figure 6: Center positions of our pointings in an Aitoff projection of Mars. The sizes of the circles correlate with the absolute value of the corrected RVs of the Martian $CO_2$ profiles, i.e. ranging from |0.001| to |0.182| km s$^{-1}$.

During the first two visits, there are some hints of possible strong winds that may be linked to dust storms on the Martian surface in that period (Guha et al., 2024; Sánchez-Lavega et al., 2025): even if slightly overestimated, the RVs values agree within the errors with the expected strength of the Martian horizontal winds, that are typically about some tens of m s$^{-1}$ at maximum (Holstein-Rathlou et al., 2010; Stott et al., 2023). The third visit shows much greater variability, with large outlier values (greater than 90 m s$^{-1}$ in absolute value). We note though that there is a higher scattering between the RVs of the solar and $CO_2$ components (see right panel in Fig. 4), so that our correction for that visit may be suboptimal. While on Earth the vertical winds may reach up to 100 m s$^{-1}$ and more (Smith, 1998), it is highly improbable that the same values can be reached on Mars, so we cannot reliably identify the cause of our RVs variations during the third visit.

### 3.2. Equivalent widths of the $CO_2$ profiles

Taking advantage of the high SNR ratio of our data, we decided to investigate relative variations in the abundances of $CO_2$ using the equivalent width (EW) of the $CO_2$ mean line profiles as a proxy. The EWs are commonly used to compute stellar abundances using the individual lines of an element (e.g. Sneden, 1973; Magrini et al., 2022; Biazzo et al., 2022): in our case we are not interested in an absolute value of the $CO_2$ abundance, already well-known in the case of Mars, but only in the variation of the signal. Therefore we can compare directly the EW values to check for any hint of variability.

We computed the EWs of the normalised profiles using two different methods, to ensure that our results were not dependent on the computational method. First, we used the same Gaussian fit performed to compute the RVs, the EW is thus the area of the Gaussian, as given by Eq. 1:

$$EW = \sqrt{2\pi} \times \sigma \times d \quad (1)$$

where *d* is the depth of the Gaussian.

Alternatively, we computed the EW as the zero moment of the line profiles (Briquet and Aerts, 2003), i.e. as the area of the observed data instead of that of the fitting function. We still used the Gaussian fit to automatically define the line limits as the center of the line plus and minus 3 σ. The two datasets agree quite well (with a Spearman correlation value of 0.998, and a p-value of $1.6 \times 10^{-20}$), aside from a small zero-point shift of 0.03, that may depend on the different continuum definition between the two methods, and we recovered the same variations with both, as can be observed by comparing the results from the Gaussian fitting (upper panel of Fig. 7) with those from the moment method (lower panel of Fig. 7).

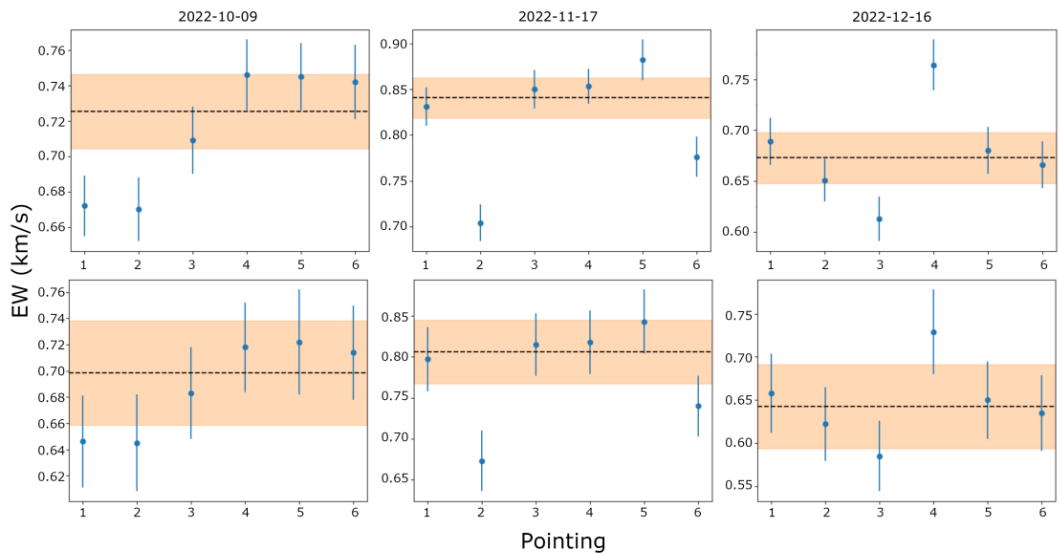

Figure 7: Spatial variations of EW of the Martian $CO_2$ profiles. *Upper panel:* results obtained computing the EW with Gaussian fitting. The dashed black line is the median value, the shaded orange area shows the region around the median within the maximum error computed on the measured EWs. *Lower panel:* as above, but using the results obtained with the moment method.

The errors on the zero moment are much larger than those derived from the Gaussian fitting, because they are estimated from the standard deviation on the continuum outside the line, and we have a heavy distortion of the continuum due to the correlation with the $CO_2$ lines excluded from our mask to avoid solar contamination and to a known fringing effect present in the GIANO-B spectra (Rainer, 2019). Thus we selected the Gaussian fitting EWs as our final results (see Table 3). Figure 8 shows the EWs values on an Aitoff projection of Mars. The second visit shows consistently higher EWs values: we excluded an incorrect telluric removal as the cause by comparing the atmospheric parameters found by `molecfit` in the three nights and finding them consistent. We also searched for correlation between the EWs and the telluric atmospheric condition during the observations (airmass, humidity, pressure, wind speed, temperature), but we found none (the Spearman correlation between EWs and the above values yielded p-values ranging from 0.47 and 0.60), so, as in the case of the RV variations, we can assume that the EW variations found are real and neither of telluric origin nor an artifact of the reduction process, which is the same for each spectrum.

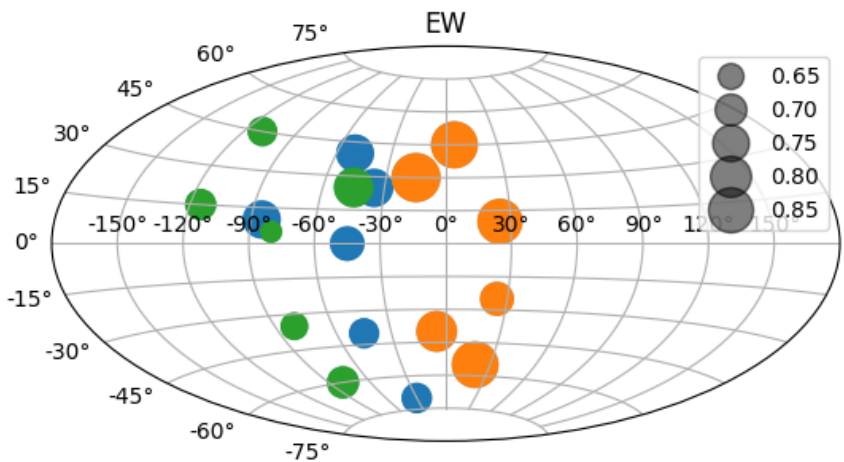

Figure 8: Center positions of our pointings in an Aitoff projection of Mars. The sizes of the circles correlate with the EWs of the Martian $CO_2$ profiles. The 2022-10-09 pointings are shown in blue, the 2022-11-17 ones in orange, and the 2022-12-16 ones in green.

We computed the $CO_2$ mean line profiles using the same mask on homogeneously normalised, telluric-corrected spectra: we can interpret the variations on the EWs of our profiles as a signal of the different topographical altitudes in those regions: the higher the altitude, the thinner the atmospheric layer we are sampling, and thus we could expect larger EWs in correspondence of plains and lower EWs in correspondence of highlands or mountains.

To explore this possible correlation, we recovered the surface altitude along the slit from the pointing positions (see Appendix A): we list the average altitude and its standard deviation in Table 1. The correlation between the measured EWs and the average altitude is shown in Fig. 9, both for all the pointings and for the average values in four bins of EW. We used the `python` module `scipy` to compute the p-value of our distribution, and found it to be 0.002 using the Spearman statistics: the p-value is below the 3 significance level, denoting that the observed correlation is unlikely due to random chance.

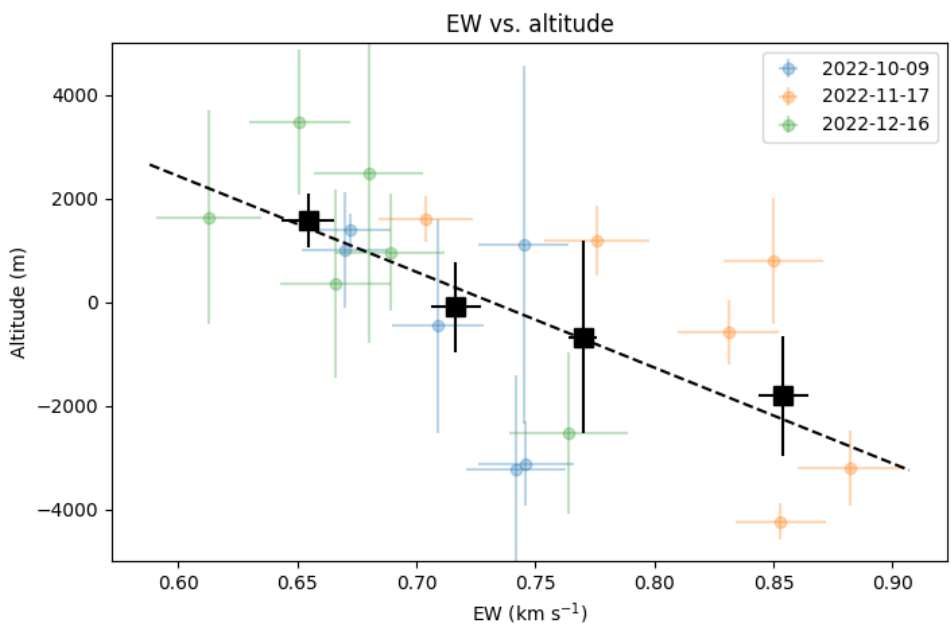

Figure 9: Comparison between the EWs of the $CO_2$ profiles (x-axis) and the average altitude of the pointing (y-axis). The altitude error bars refer to the standard deviation of the altitude sampled by each pointing. The black squares show the average values in four EW ranges ([0.600:0.675] [0.675:0.750] [0.750:0.825] [0.825:0.900]). The dashed line show the linear correlation between EWs and altitudes.

Similarly to what we did in our analysis of RVs variations, also for the study of the EWs variations we could use the solar mean line profiles as a comparison: we compared the solar EWs with the sampled altitudes (see Fig. 10) and we found no correlation in the data (p-value=0.27 with the Spearman statistics). The two simultaneous datasets (Martian $CO_2$ profiles and solar mean line profiles) behave differently, so we can reasonably exclude reduction artifacts or instrumental effects as a cause of the Martian EWs variability.

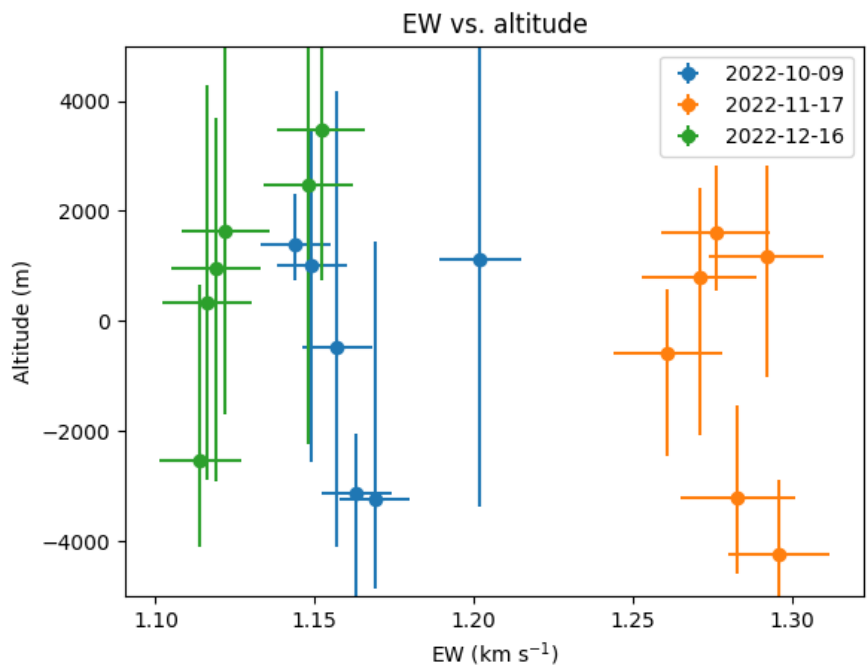

Figure 10: Comparison between the EWs of the solar profiles (x-axis) and the average altitude of the pointing (y-axis).

**4. Results**

We found variability both in the RVs and the EWs of the Martian $CO_2$ profiles. While the mean RV value is zero for each visit and also for all the visits combined, there is a large variability between the pointings, that may be due to strong winds on the planet's surface. The EW variations correlate in a statistically significant way with the average altitude of the regions observed within GIANO-B's slit: the lower the altitude of the observed Martian region, the larger the atmospheric layers sampled and thus the greater the EWs. As Bibring et al. (1991) used low-resolution spectroscopic measurements of $CO_2$ column density to recover the topology of Mars, so we followed the inverse path and used the well-known Martian topography to explain our high-resolution EW measurements.

While our pointings cover quite large regions on the Martian surface (~2500-3000 km s$^{-1}$ along the slit's length), we are still able to look at finer details than what we would observe in exoplanetary observations in reflected light. In fact, in the latter case we could not resolve the exoplanet's surface at all and we would observe always a whole hemisphere: as such, any atmospheric signal would only sample the global properties of the visible hemisphere. Because no telluric planet is expected to be perfectly homogeneous topographically, we would still be able to infer some global scale geological properties from the variations of the atmospheric signal observed during different visits, in the case of thin Martian-like atmospheres. Thick Venus-like cloud coverages or Earth-like atmospheres would either hinder the observations of surface topography or be dominated by atmospheric effects (e.g., Fujii at al., 2013).

The best way to have Martian observations of very different global characteristics would have been the sampling of the two hemispheres (i.e. lowlands and highlands) divided by the Martian dichotomy (Watters et al., 2007). Unfortunately, due to the inclination of Mars during our visits we were not able to study separately the northern lowlands and the southern highlands to highlight possible variations related to the Martian dichotomy. In fact, our three uppermost northern pointings lie between 30° and 45°, at the upper boundary between the two hemispheres. Still, we could consider the global view of Mars during our visits: while the first visit (2022-10-09) partially overlaps with the other two, the second (2022-11-17) and the third (2022-12-16) ones explore different faces of the planet. We averaged our data to mimic an observation of the whole planetary face, and we found a statistically significant variation between the second and the third visit, with an average EW of 0.82±0.02 km s$^{-1}$ (2022-11-17) versus 0.68±0.02 km s$^{-1}$ (2022-12-16). This result correlates also with the average altitude of the two visits, i.e. -745±910 m in 2022-11-17 versus 1055±775 m in 2022-12-16: also in this case, the larger EW is linked to a lower altitude and thus to a thicker atmospheric layer probed.

While we are unable to reach the same level of SNR in exoplanets data with the current ground-based high-resolution spectrograph, we can compare our results with the expected ones from ANDES (Palle et al., 2025). Exploiting the vast collecting area of the ELT, ANDES will be able to see extremely faint signals: it is estimated that it would be possible to recover the atmospheric signature of $H_2O$ in Proxima Centauri b in reflected light with the CCF technique in a single night of observation, in the best case scenario (thin 0.1 bar Earth-like atmosphere with clouds). A $CO_2$ dominated atmosphere would require more nights of observations, and thus it would be difficult to link variations in the signal to topographical effects. In any case, if we degrade our LSD profiles adding random Gaussian noise (see Fig. 11) in order to reach a signal around the 5 σ threshold used in Palle et al. (2025), we still recover a 1 σ difference between the EWs of the most different pointing (pointing 5 of 2022-11-17 and pointing 3 of 2022-12-16, respectively the rightmost and leftmost points in Fig. 9).

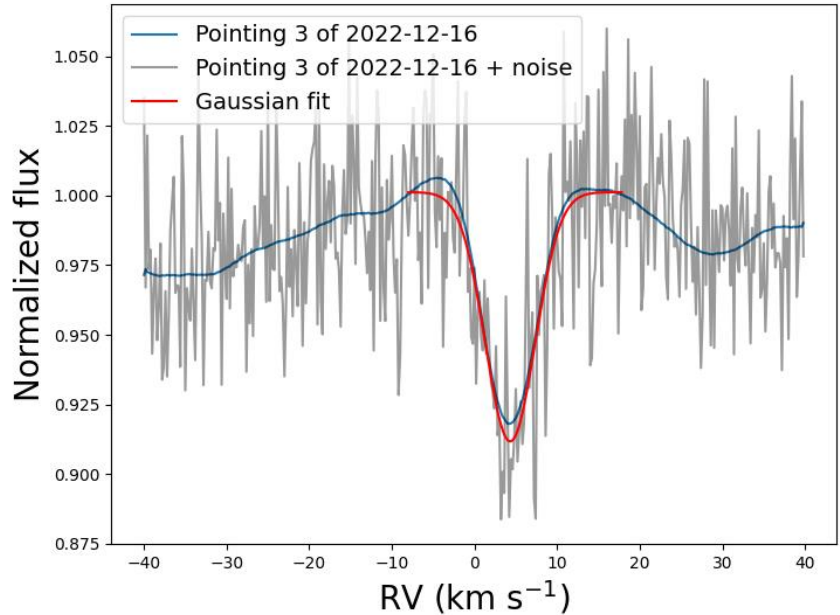

Figure 11: Original (blue) and degraded (grey) LSD profile of pointing 3 of 2022-12-16. The Gaussian fit of the degraded profile is plotted in red.

We note here that ANDES will have a resolution of about 100,000 versus the 50,000 of GIANO-B, and this will help to identify variations in the EWs of the CCF profile, that will be built with a larger number of unblended absorption lines. Also, we used a simple Gaussian fitting to recover both RVs and EWs, while more sophisticated analysis may help lower the errors on the results. Additionally, summing more nights of observations will of course increase the SNR of the recovered signals, but this would imply adding data from different views of the planetary surfaces. If the rotation period of the exoplanet could be recovered, then it would be possible to schedule observations of the same hemisphere and then combine them, but up to now the rotation period of exoplanets has been computed only for a handful of cases (e.g., Snellen et al., 2014).

## 5. Conclusions and future works

We used Mars as a Solar System analog for a telluric exoplanet, to verify the possibility of linking variations in the atmospheric signal computed with exoplanetary techniques to the geological setting of the planet. We used the very well-known Martian topography as a reference guide to interpret the variability of the atmospheric signal of telluric exoplanets observed with high-resolution spectroscopy in reflected light. We planned to study both temporal (seasonal) and spatial variations of the Martian atmospheric $CO_2$ signal, but unfortunately we were unable to do the former due to the bad weather conditions affecting our observations. We were limited to three visits during a three-month period, all of them before the Martian equinox.

Using the images of the GIANO-B guide we were able to identify the position of our pointings on the Martian surface and then recover information on the average altitude of the surface along the slit. We found variations in the RVs of the $CO_2$ signal that may be caused by the presence of winds, with a median speed of 0±71 m s$^{-1}$, i.e., the mean value of zero is consistent with the Martian reference frame, while the spread is due to the RV variations in the different pointings. We found an inverse correlation between the strength of the $CO_2$ signal, measured by its EW, and the altitude, that is consistent with stronger signals being caused by probing larger atmospheric layers.

To better mimic the signal recovered by exoplanetary studies, we averaged the signal obtained during the second (2022-11-17) and third (2022-12-16), which have almost no spatial overlapping. This is similar to observing different hemispheres during observations performed in different nights in reflected light of non tidally locked telluric exoplanets. We found a statistically significant difference in the strength of the signal during the two visits, and we identified its cause as the different average altitude of the Martian faces seen in the two visits.

We demonstrated here that it is possible, with high-resolution spectra with a wide wavelength range and high SNR, to obtain information on a planet's topography in the case of a thin, Martian-like atmosphere: the different strengths of the exoplanetary atmospheric signal in reflected light during different observing nights could be an indicator of the different regional physiographical features of the reflecting hemisphere (e.g. northern lowlands and southern high-lands), and as such they could indicate the existence of significant topographical variability at the surface.

On Earth, it is widely known that strong topographic contrasts are often related to geodynamic processes and tectonic activity (Burbank and Anderson, 2013). Similar evidence has been identified also on other planets of the Solar System such as Mars (Wenzel et al., 2004; Hauber et al., 2010; Rossi and van Gasselt, 2010; Balbi et al., 2024). In this way, inferring the existence of physiographical features on exoplanetary surfaces could represent the first step to understand if geodynamic processes have been active on the investigated exoplanet.

In the future, we plan to investigate the variability of different atmospheric molecules (e.g. $CH_4$), exploiting the large wavelength range and the good quality of our spectra. We also plan to modify the `Gofio` program in order to extract the spectra in different slices of the slit, to allow for a more precise investigation of the link between the atmospheric signal and the spatial geographical and geophysical characteristics of the observed regions: this could allow us to exploit our data for a more detailed study of Mars itself, and not only as a telluric exoplanet analog.

*Acknowledgements*

We thank the TNG staff, who helped us to carry on a non-standard observation plan, and who worked beyond expectations to ensure that we received as much data as possible despite the unfortunate weather conditions that we encountered.

**Appendix A. Pointings**

We show here the North-East alignment (Fig. A.12), the pointings (Figs. A.13, A.14, A.15), and the slit's drift (Figs. A.16, A.17, and A.18) for all the three visits we performed on Mars.

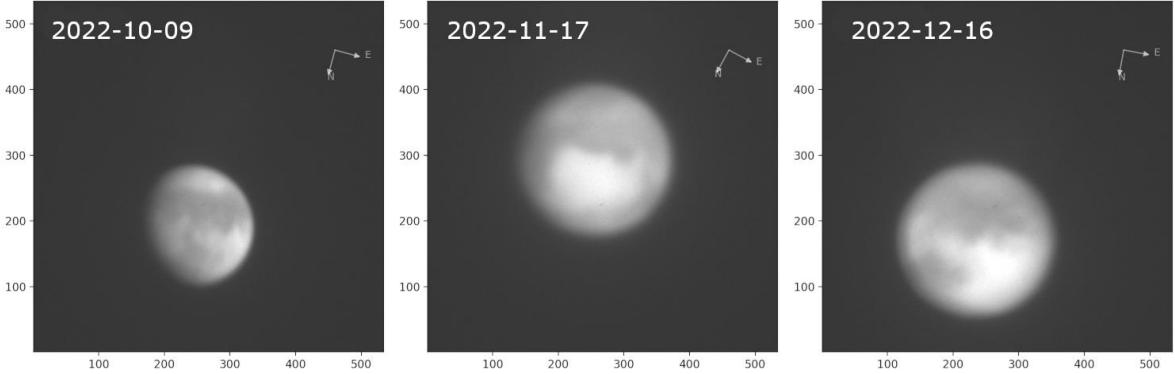

Figure A.12: Alignment of Mars during the three visits.

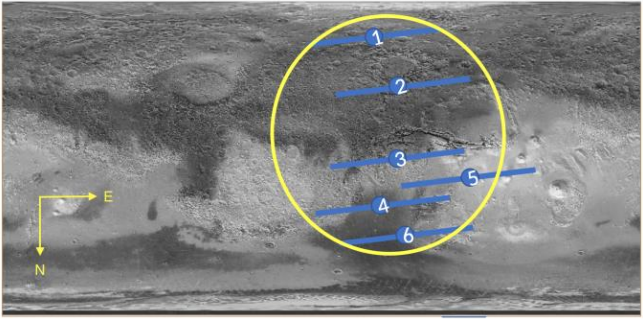

Figure A.13: Pointings during the 2022-10-09 visit. To better compare our identification with the slit's positions, we aligned the map in same configuration of the pointings.

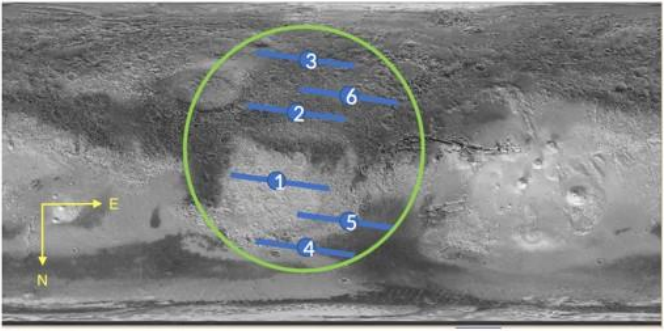

Figure A.14: Pointings during the 2022-11-17 visit. To better compare our identification with the slit's positions, we aligned the map in same configuration of the pointings.

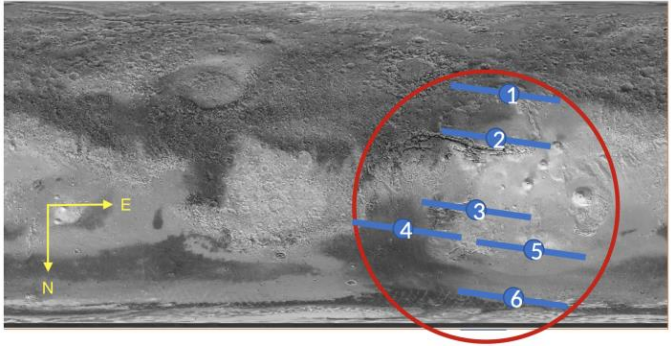

Figure A.15: Pointings during the 2022-12-16 visit. To better compare our identification with the slit's positions, we aligned the map in same configuration of the pointings.

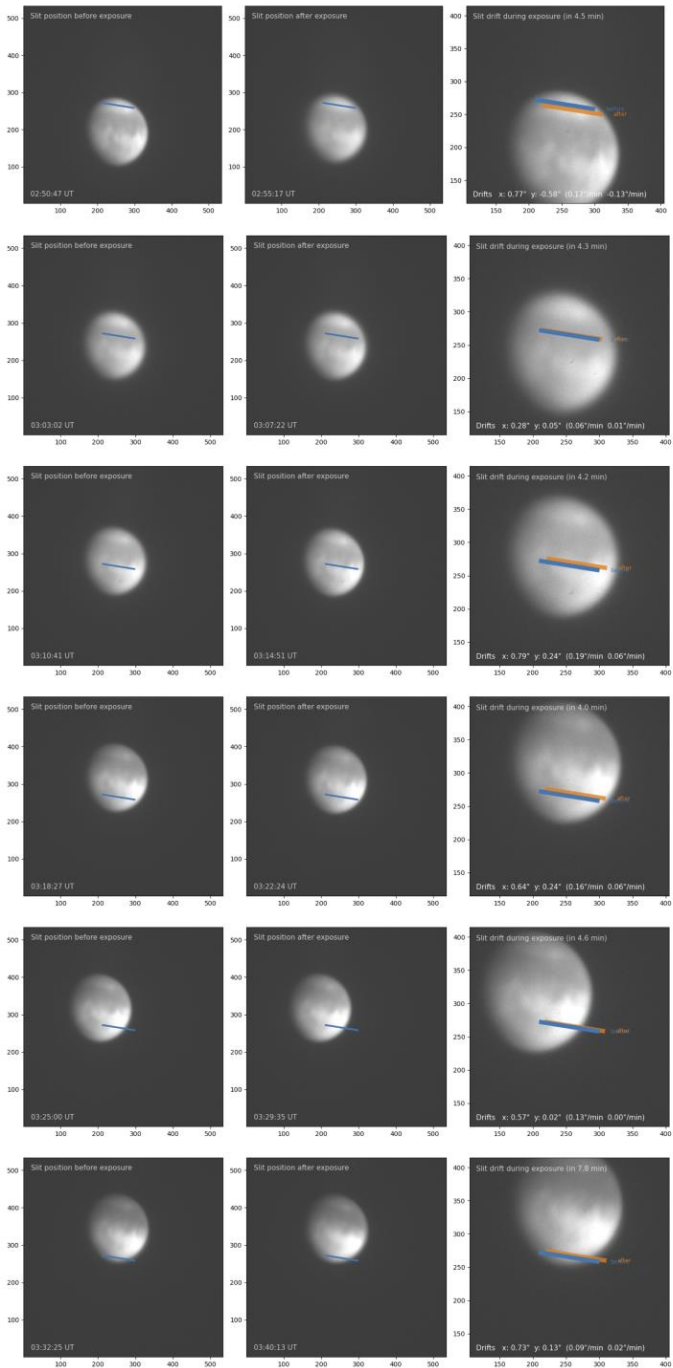

Figure A.16: Pointings and slit's drift during the 2022-10-09 visit.

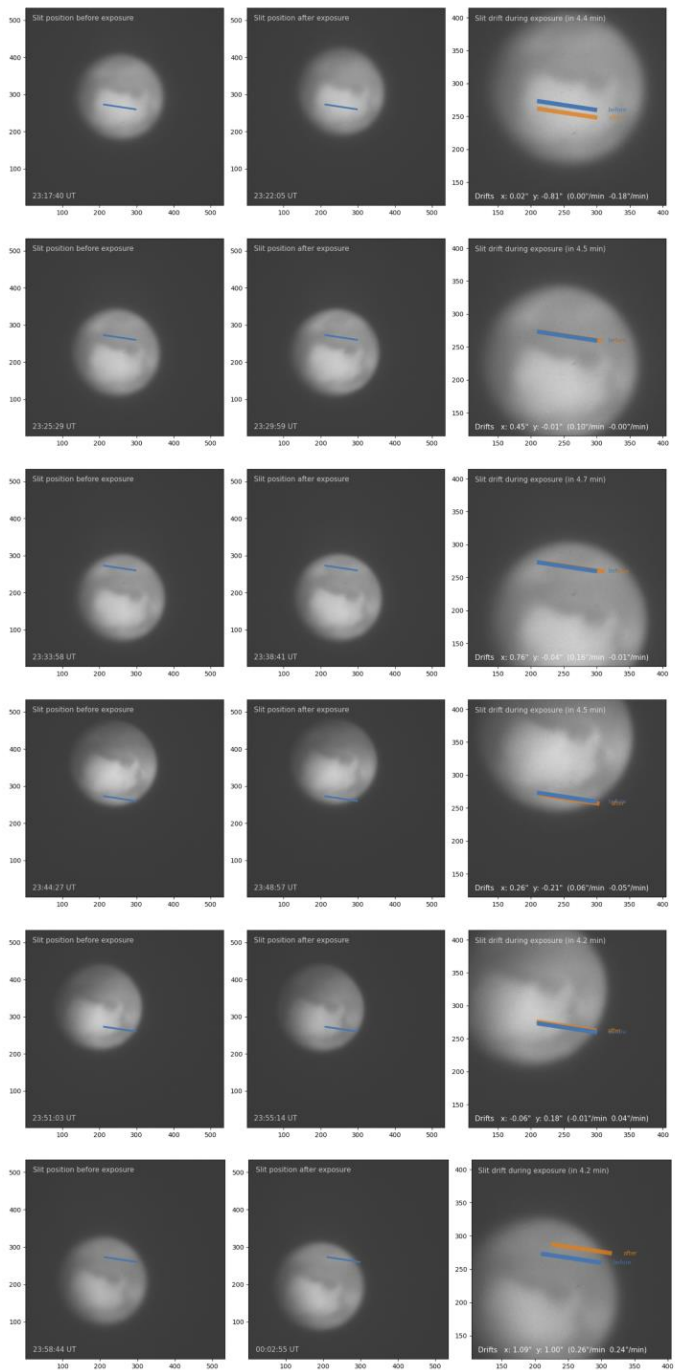

Figure A.17: Pointings and slit's drift during the 2022-11-17 visit.

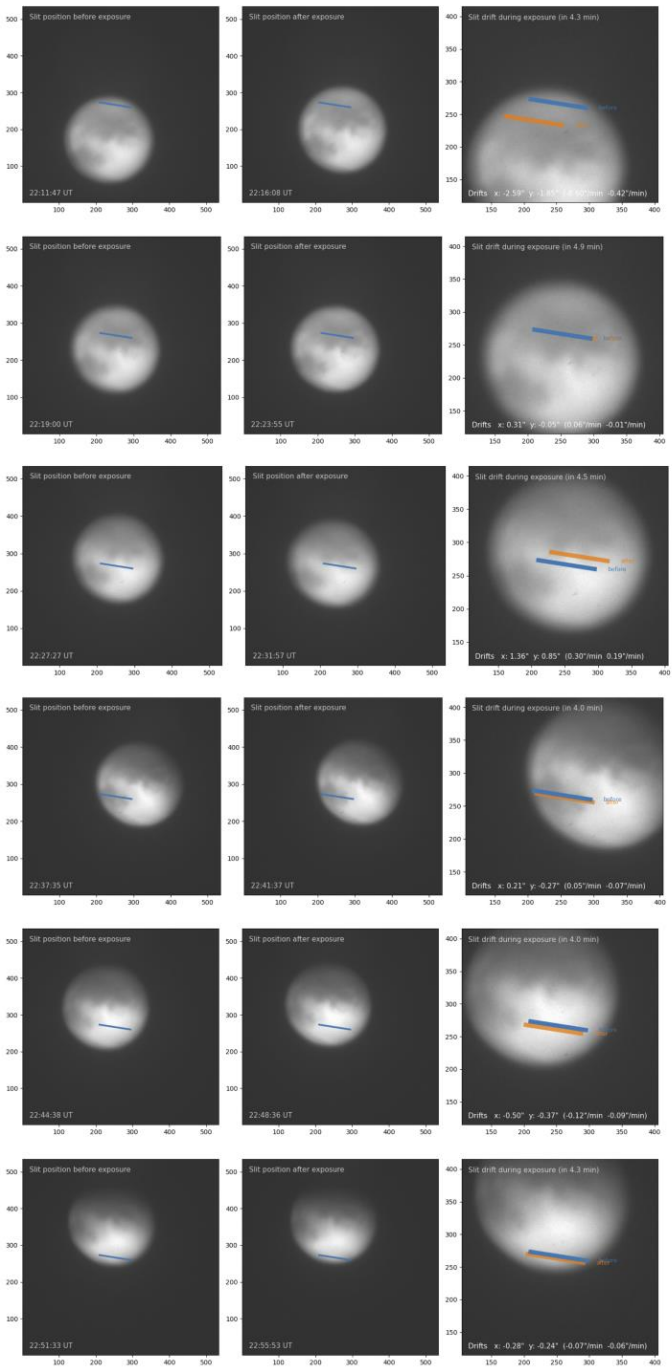

Figure A.18: Pointings and slit's drift during the 2022-12-16 visit.

**References**

Asplund, M., Amarsi, A.M., Grevesse, N., 2021. The chemical make-up of the Sun: A 2020 vision. A&A 653, A141, doi: https://doi.org/10.1051/0004-6361/202140445

Balbi, E., Ferretti, G., Tosi, S., Crispini, L., Cianfarra, P., 2024. Polyphase tectonics on mars: Insight from the claritas fossae. Icarus 411, 115972, doi: https://doi.org/10.1016/j.icarus.2024.115972

Biazzo, K., et al., 2022. The GAPS Programme at TNG. XXXV. Fundamental properties of transiting exoplanet host stars. A&A 664, A161, doi: https://doi.org/10.1051/0004-6361/202243467

Bibring, J.P., et al., 1991. Topography of the Martian tropical regions with ISM. P&SS 39, 225–236, doi: https://doi.org/10.1016/0032-0633(91)90146-2

Birkby, J.L., 2018.Spectroscopic Direct Detection of Exoplanets, in: Deeg, H.J., Belmonte, J.A. (Eds.), Handbook of Exoplanets, p. 16, doi: https://doi.org/10.1007/978-3-319-55333-7_16

Birkby, J.L., de Kok, R.J., Brogi, M., Schwarz, H., Snellen, I.A.G., 2017. Discovery of Water at High Spectral Resolution in the Atmosphere of 51 Peg b. AJ 153, 138, doi: https://doi.org/10.3847/1538-3881/aa5c87

Borsa, F., et al., 2021a. Atmospheric Rossiter-McLaughlin effect and transmission spectroscopy of WASP-121b with ESPRESSO. A&A 645, A24, doi: https://doi.org/10.1051/0004-6361/202039344

Borsa, F., et al., 2021b. The GAPS Programme at TNG. XXXI. The WASP-33 system revisited with HARPS-N. A&A 653, A104, doi: https://doi.org/10.1051/0004-6361/202142768

Borsa, F., et al., 2022. The GAPS Programme at TNG. XXXIII. HARPS-N detects multiple atomic species in emission from the dayside of KELT-20b. A&A 663, A141, doi: https://doi.org/10.1051/0004-6361/202140559

Briquet, M., Aerts, C., 2003. A new version of the moment method, optimized for mode identification in multiperiodic stars. A&A 398, 687–696, doi: https://doi.org/10.1051/567 0004-6361:20021683

Brogi, M., et al., 2012. The signature of orbital motion from the dayside of the planet τ Boötis b. Nature 486, 502–504, doi: https://doi.org/10.1038/nature11161

Burbank, D.W., Anderson, R.S., 2013. Tectonic Geomorphology, Second Edition. Environmental & Engineering Geoscience 19, 198–200, doi: https://doi.org/10.2113/gseegeosci.19.2.198

Carleo, I., et al., 2022. The GAPS Programme at TNG XXXIX. Multiple Molecular Species in the Atmosphere of the Warm Giant Planet WASP-80 b Unveiled at High Resolution with GIANO-B. AJ 164, 101, doi: https://doi.org/10.3847/1538-3881/ac80bf

Christensen, P.R., et al., 2001. Mars Global Surveyor Thermal Emission Spectrometer experiment: Investigation description and surface science results. JGR 106, 23823–23872, doi: https://doi.org/10.1029/2000JE001370

Claudi, R., et al., 2018. GIARPS: commissioning and first scientific results, in: Evans, C.J., Simard, L., Takami, H. (Eds.), Ground-based and Airborne Instrumentation for Astronomy VII, p. 107020Z, doi: https://doi.org/10.1117/12.2312555

Demory, B.-O., et al., 2016. A map of the large day-night temperature gradient of a super-Earth exoplanet, Nature 532, 7598, doi: https://doi.org/10.1038/nature17169

Di Marcantonio, P., Morossi, C., Franchini, M., Lehmann, H., 2019. Using Independent Component Analysis to Detect Exoplanet Reflection Spectrum from Composite Spectra of Exoplanetary Binary Systems. AJ 158, 161, doi: https://doi.org/10.3847/1538-3881/ab3e71

Donati, J.F., Semel, M., Carter, B.D., Rees, D.E., Collier Cameron, A., 1997. Spectropolarimetric observations of active stars. MNRAS 291, 658–682, doi: https://doi.org/10.1093/mnras/291.4.658

Fergason, R., Hare, T., Laura, J., 2018. Hrsc and mola blended digital elevation model at 200m v2. astrogeology pds annex. US Geological Survey 6

Franz, H.B., et al., 2017. Initial SAM calibration gas experiments on Mars: Quadrupole mass spectrometer results and implications. P&SS 138, 44–54, doi: https://doi.org/10.1016/j.pss.2017.01.014

Fujii, Y., et al., 2013, Variability of Water and Oxygen Absorption Bands in the Disk-integrated Spectra of Earth, ApJ 765, 76, doi: https://doi.org/10.1088/0004-637X/765/2/76

Giacobbe, P., et al., 2021. Five carbon- and nitrogen-bearing species in a hot giant planet's atmosphere. Nature 592, 205–208, doi: https://doi.org/10.1038/s41586-021-03381-x

Guha, B.K., et al., 2024, Seasonal and Diurnal Variations of Dust Storms in Martian Year 36 Based on the EMM-EXI Database, JGR Planets 129, 4, doi: https://doi.org/10.1029/2023JE008156

Guilluy, G., et al., 2022. The GAPS Programme at TNG. XXXVIII. Five molecules in the atmosphere of the warm giant planet WASP-69b detected at high spectral resolution. A&A 665, A104, doi: https://doi.org/10.1051/0004-6361/202243854

Haberle, R., 2015. Solar system/sun, atmospheres, evolution of atmospheres — planetary atmospheres: Mars, in: North, G.R., Pyle, J., Zhang, F. (Eds.), Encyclopedia of Atmospheric Sciences (Second Edition). second edition ed. Academic Press, Oxford, pp. 168–177, doi: https://doi.org/10.1016/B978-0-12-382225-3.00312-1

Harutyunyan, A., et al., 2018. GIANO-B online data reduction software at the TNG, in: Navarro, R., Geyl, R. (Eds.), Advances in Optical and Mechanical Technologies for Telescopes and Instrumentation III, p. 1070642, doi: https://doi.org/10.1117/12.2312690

Hauber, E., Grott, M., Kronberg, P., 2010. Martian rifts: Structural geology and geophysics. Earth and Planetary Science Letters 294, 393–410, doi: https://doi.org/10.1016/j.epsl.2009.11.005

Holstein-Rathlou, C., et al., 2010. Winds at the Phoenix landing site. Journal of Geophysical Research (Planets) 115, E00E18, doi: https://doi.org/10.1029/2009JE003411

Kausch, W., et al., 2015. Molecfit: A general tool for telluric absorption correction. II. Quantitative evaluation on ESO-VLT/X-Shooter spectra. A&A 576, A78, doi: https://doi.org/10.1051/0004-6361/201423909

Kitzmann, D., et al., 2023. The Mantis network. A standard grid of templates and masks for cross-correlation analyses of ultra-hot Jupiter transmission spectra. A&A 669, A113, doi: https://doi.org/10.1051/0004-6361/202142969

Kjeldsen, H., Bedding, T.R., 2011. Amplitudes of solar-like oscillations: a new scaling relation. A&A 529, L8, doi: https://doi.org/10.1051/0004-6361/201116789

Kochukhov, O., Makaganiuk, V., Piskunov, N., 2010. Least-squares deconvolution of the stellar intensity and polarization spectra .A&A 524, A5, doi: https://doi.org/10.1051/0004-6361/201015429

Lanza, A.F., Molaro, P., Monaco, L., Haywood, R.D., 2016. Long-term radial velocity variations of the Sun as a star: The HARPS view. A&A 587, A103, doi: https://doi.org/10.1051/0004-6361/201527379

Loftus, K., et al., 2025. Extreme weather variability on hot rocky exoplanet 55 Cancri e explained by magma temperature-cloud feedback. PNAS, 122(17), e2423473122, doi: https://doi.org/10.1073/pnas.2423473122

Magrini, L., et al., 2022. Ariel stellar characterisation. I. Homogeneous stellar parameters of 187 FGK planet host stars: Description and validation of the method. A&A 663, A161, doi: https://doi.org/10.1051/0004-6361/202243405

Marconi, A., et al., 2022. ANDES, the high resolution spectrograph for the ELT: science case, baseline design and path to construction, in: Evans, C.J., Bryant, J.J., Motohara, K. (Eds.), Ground-based and Airborne Instrumentation for Astronomy IX, p. 1218424, doi: https://doi.org/10.1117/12.2628689

Mayor, M., Queloz, D., 1995. A Jupiter-mass companion to a solar-type star. Nature 378, 355–359, doi: https://doi.org/10.1038/378355a0

Meftah, M., Sarkissian, A., Keckhut, P., Hauchecorne, A., 2023. The SOLAR HRS New High-Resolution Solar Spectra for Disk-Integrated, Disk-Center, and Intermediate Cases. Remote Sensing 15, 3560, doi: https://doi.org/10.3390/rs15143560

Molliere, P., et al., 2019. petitRADTRANS. A Python radiative transfer package for exoplanet characterization and retrieval. A&A 627, A67, doi: https://doi.org/10.1051/0004-6361/201935470

Nugroho, S.K., et al., 2020. Searching for thermal inversion agents in the transmission spectrum of KELT-20b/MASCARA-2b: detection of neutral iron and ionised calcium H&K lines. MNRAS 496, 504–522, doi: https://doi.org/10.1093/mnras/staa1459

Oliva, E., et al., 2012. The GIANO spectrometer: towards its first light at the TNG, in: McLean, I.S., Ramsay, S.K., Takami, H. (Eds.), Ground-based and Airborne Instrumentation for Astronomy IV, p. 84463T, doi: https://doi.org/10.1117/12.925274

Oliva, E., et al., 2018. GIANO, the high resolution IR spectrograph of the TNG: geometry of the echellogram and strategies for the 2D reduction of the spectra, in: Evans, C.J., Simard, L., Takami,

H. (Eds.), Ground-based and Airborne Instrumentation for Astronomy VII, p. 1070274, doi: https://doi.org/10.1117/12.2309927

Origlia, L., et al., 2014. High resolution near IR spectroscopy with GIANO-TNG, in: Ramsay, S.K., McLean, I.S., Takami, H. (Eds.), Ground-based and Airborne Instrumentation for Astronomy V, p. 91471E, doi: https://doi.org/10.1117/12.2054743

Palle, E., et al., 2025. Ground-breaking Exoplanet Science with the ANDES spectrograph at the ELT. ExA 59, 29, doi: https://doi.org/10.1007/s10686-025-10000-4

Parmentier, V., et al., 2018. From thermal dissociation to condensation in the atmospheres of ultra hot Jupiters: WASP-121b in context. A&A 617, A110, doi: https://doi.org/10.1051/0004-6361/201833059

Patel, J. A., et al., 2024. JWST reveals the rapid and strong day-side variability of 55 Cancri e. A&, 690, A159, doi: https://doi.org/10.1007/s10686-025-10000-4

Pino, L., et al., 2020. Neutral Iron Emission Lines from the Dayside of KELT-9b: The GAPS Program with HARPS-N at TNG XX. ApJ 894, L27, doi: https://doi.org/10.3847/2041-8213/ab8c44

Queloz, D., 1995, Echelle Spectroscopy with a CCD at Low Signal-To-Noise Ratio, IAUS 167, 221

Rainer, M., 2019. GIANO-B reduction pipeline: GOFIO manual, doi: https://doi.org/10.20371/INAF/SW/2019_00002 URL: http://hdl.handle.net/20.500.12386/33712

Rainer, M., et al, 2021. The GAPS programme at TNG. XXX. Atmospheric Rossiter-McLaughlin effect and atmospheric dynamics of KELT-20b. A&A 649, A29, doi: https://doi.org/10.1051/0004-6361/202039247

Rainer, M., et al., 2018. Introducing GOFIO: a DRS for the GIANO-B near-infrared spectrograph, in: Evans, C.J., Simard, L., Takami, H. (Eds.), Ground-based and Airborne Instrumentation for Astronomy VII, p. 1070266, doi: https://doi.org/10.1117/12.2312130

Rossi, A.P., van Gasselt, S., 2010. Geology of mars after the first 40 years of exploration. Research in Astronomy and Astrophysics 10, 621, doi: https://doi.org/10.1088/1674-4527/10/7/003

Sánchez-Lavega, A., et al., 2025, Martian Atmospheric Disturbances From Orbital Images and Surface Pressure at Jezero Crater, Mars, During Martian Year 36, JGR Planets 130, 1, doi: https://doi.org/10.1029/2024JE008565

Smette, A., et al., 2015. Molecfit: A general tool for telluric absorption correction. I. Method and application to ESO instruments. A&A 576, A77, doi: https://doi.org/10.1051/0004-6361/201423932

Smith, R.W., 1998. Vertical winds: a tutorial. Journal of Atmospheric and Solar-Terrestrial Physics 60, 1425–1434, doi: https://doi.org/10.1016/S1364-6826(98)00058-3

Sneden, C., 1973. The nitrogen abundance of the very metal-poor star HD 122563. ApJ 184, 839, doi: https://doi.org/10.1086/152374

Snellen, I.A.G., de Kok, R.J., de Mooij, E.J.W., Albrecht, S., 2010. The orbital motion, absolute mass and high-altitude winds of exoplanet HD209458b. Nature 465, 1049–1051, doi: https://doi.org/10.1038/nature09111

Snellen. I.A.G., et al., 2014, Fast spin of the young extrasolar planet β Pictoris b, Nature 509, 63, doi: https://doi.org/10.1038/nature13253

Stott, A.E., et al., 2023. Wind and Turbulence Observations with the Mars Microphone on Perseverance. Journal of Geophysical Research (Planets) 128, e2022JE007547, doi: https://doi.org/10.1029/2022JE007547

Strachan, J.B.P., Anglada-Escude, G., 2020. Doppler shifts and spectral line profile changes in the starlight scattered from an exoplanet. MNRAS 493, 1596–1613, doi: https://doi.org/10.1093/mnras/staa268

Wang, J., Fischer, D.A., Horch, E.P., Huang, X., 2015. On the Occurrence Rate of Hot Jupiters in Different Stellar Environments. ApJ 799, 229, doi: https://doi.org/10.1088/0004-637X/799/2/229

Watters, T.R., McGovern, P.J., Irwin III, R.P., 2007. Hemispheres apart: The crustal dichotomy on mars. Annual Review of Earth and Planetary Sciences 35, 621–652, doi: https://doi.org/10.1146/annurev.earth.35.031306.140220

Wenzel, M.J., Manga, M., Jellinek, A.M., 2004. Tharsis as a consequence of mars' dichotomy and layered mantle. Geophysical Research Letters 31, doi: https://doi.org/10.1029/2003GL019306

Wyttenbach, A., Ehrenreich, D., Lovis, C., Udry, S., Pepe, F., 2015. Spectrally resolved detection of sodium in the atmosphere of HD 189733b with the HARPS spectrograph. A&A 577, A62, doi: https://doi.org/10.1051/0004-6361/201525729

Yan, F., Henning, T., 2018. An extended hydrogen envelope of the extremely hot giant exoplanet KELT-9b. Nature Astronomy 2, 714–718, doi: https://doi.org/10.1038/s41550-018-0503-3